\documentclass[epj]{svjour}
\usepackage{graphics}
\usepackage{textcomp}
\usepackage{bm}
\usepackage{amssymb}
\usepackage{amsmath}
\begin{document}
\title{Measurement of the trapping lifetime close to a cold metallic surface on a cryogenic atom-chip}
\author{Andreas Emmert\inst{1} \and Adrian Lupa{\c s}cu\inst{1} \and Gilles
Nogues\inst{1} \and Michel Brune\inst{1} \and Jean-Michel Raimond\inst{1} \and
Serge Haroche\inst{1,}\inst{2}
}                     
\offprints{G. N., \texttt{gilles.nogues@lkb.ens.fr}}

\institute{Laboratoire Kastler Brossel, ENS, CNRS, UPMC, 24
 rue Lhomond F-75231 Paris cedex 05\and Coll\`ege de France, 11 place Marcelin
Berthelot F-75231 Paris cedex 05}
\date{Received: date / Revised version: date}
%
\abstract{
We have measured the trapping lifetime of magnetically trapped atoms in a cryogenic atom-chip experiment. An ultracold atomic cloud is kept at a fixed distance from a thin gold layer deposited on top of a superconducting trapping wire. The lifetime is studied as a function of the distances to the surface and to the wire. Different regimes are observed, where loss rate is determined either by the technical current noise in the wire or the Johnson-Nyquist noise in the metallic gold layer, in good agreement with theoretical predictions. Far from the surface, we observe exceptionally long trapping times for an atom-chip, in the 10-minutes range.
} 
\authorrunning{A. Emmert {\sl et al.}}
\titlerunning{Measurement of the trapping lifetime}
\maketitle
\section{Introduction}
\label{sec:intro}
In atom-chip experiments, cold atoms are trapped in magnetic field gradients created by microfabricated current-carrying wires~\cite{TR_ZIMMERMANRMPCHIP07} or ferromagnetic structures~\cite{TR_HINDSREVIEWFERROMAG99}.
Microfabrication techniques allow for the design of static or r.f. fields in the close vicinity of the atoms. By adjusting these fields, the atomic cloud can be manipulated in a very versatile manner. The large achievable oscillation frequencies in the traps pave the way to the quantum control of the atomic motion. Moreover, ultracold atoms can be integrated on the chip with other microfabricated devices like nanomechanical resonators~\cite{TR_REICHEL_NANOMECH_07}, SQUIDs~\cite{TR_SINGHBECLOOP07} or cavities~\cite{TR_REICHEL_BECCAVITY_07,TR_SCHMIEDMAYER_BECCAVSUPRA_08}. To enhance the coupling to the latter, highly excited atoms (Rydberg atoms) could be used in a cryogenic environment~\cite{ENS_RYDBERGTRAP04}.

High trapping frequencies are achieved when the atomic cloud is brought very close to the surface of the chip. In these conditions, however, new loss mechanisms are experimentally observed~\cite{TR_HINDSEXPTRAPLOSS_03,TR_VULETICCHIPNOISE04}. They originate mainly from Johnson-Nyquist noise currents in the trapping structure. They produce magnetic-field fluctuations at the position of the trapped cloud, inducing Zeeman transitions towards untrapped magnetic sublevels~\cite{TR_HINDSREKDAL_FLIPCALCULATION_04,TR_HENKELMETALNOISE05}. 

This difficulty might be overcome by operating the atom-chip at low temperature, with superconducting materials. The magnetic field noise in the vicinity of a superconductor is expected to be drastically smaller than that of a normal metal. Hence, the trapping lifetime should increase significantly~\cite{TR_HINDS_LIFETIMECONSUPERCON_05,TR_REKDALSUPERCONDNOISE06,TR_HINDSELIASHBERG07,TR_REKDAL_ELIASHBERG_07}, with interesting potentialities for coherent atom manipulations. Additionally, one benefits from an extremely good vacuum due to cryogenic pumping, as already demonstrated for magnetic traps with cen\-ti\-me\-tre-sized coils~\cite{TR_LIBBRECHTCRYOTRAP95}. These promising features have triggered experimental efforts to develop and operate cryogenic atom-chips~\cite{ENS_CHIPSUPRA06,TR_SHIMIZUPERMANENTCHIP07,ENS_BECSUPRA08}.  

Even in the vicinity of a normal metal, the use of low temperature could improve significantly the trapping lifetime. As shown in~\cite{TR_FOLMANCOLDMETALNOISE05}, the lifetime in this case scales as $[\sigma (T) \cdot T]^{-1}$, where $T$ is the temperature, proportional, down to sub-Kelvin temperatures, to the mean number of blackbody photons at the frequency of the Zeeman transitions, and $\sigma(T)$ the conductivity of the metal. When $T$ is high, $\sigma(T)$ is inversely proportional to $T$ and the lifetime is temperature-independant. For low enough temperatures, $\sigma$ saturates to a value determined by the electron diffusion on the impurities in the metal. The lifetime increases then as $1/T$. 

We report here measurements of the trapping lifetime in front of a gold surface at 4.2~K. The experiment is performed in the cryogenic setup of Refs.~\cite{ENS_CHIPSUPRA06,ENS_BECSUPRA08}. We identify two loss mechanisms. The first is due to spin-flips induced by technical current fluctuations in the trapping wire. The second is due to the Johnson-Nyquist noise in a thin gold layer of thickness $h=$~200~nm deposited on top of the superconducting niobium atom-chip. Bringing the atoms as close as 15~\textmu m to the chip along two different paths, we measure independently the loss rates associated to both mechanisms.

\section{Experimental conditions}
\label{sec:experiments}

\begin{figure}
\begin{center}
\resizebox{0.9\columnwidth}{!}{
\includegraphics{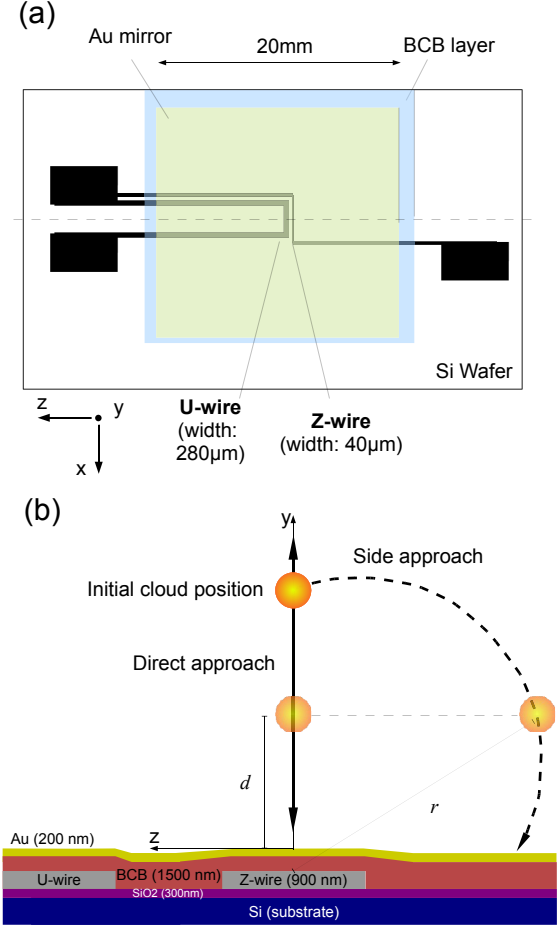}
} \end{center}
  \caption{(a) Front view of the atom-chip used in this experiment. The trapping Niobium wires (U-wire and Z-wire) are covered by an insulating BCB layer and a gold mirror. (b) Cut of the atom-chip structure in the $yz$ plane (along the dashed line in (a)). Atoms are trapped and approached to a distance $d$ from the surface, either directly over the Z-wire (direct approach, solid arrow) or by maintaining a nearly constant distance $r$ (between 160 and 240~\textmu m) to the wire center (side approach, dashed arrow).}
 \label{fig:approach}
\end{figure}

Our experimental setup is described in detail in Refs.~\cite{ENS_CHIPSUPRA06,ENS_BECSUPRA08}. The superconducting chip is cooled down to 4.2~K in a $^4$He cryostat with optical access. The current-carrying wires are etched from a Nb superconducting layer (thickness 960~nm) sputtered on a oxidized silicon substrate (Fig.~\ref{fig:approach}). The trapping structures are planarized with a 1.5~\textmu m thick layer of BCB resist. Finally, a 200~nm thick gold layer is evaporated onto the chip. It reflects the laser beams used for atomic cooling, manipulation and imaging. Centimetre-sized superconducting coils  generate a fully adjustable external bias magnetic field $\bm{B}$ in the trapping region. $^{87}$Rb atoms are  initially trapped in a 2D-MOT in a UHV cell at room temperature. They are then transferred in a slow atomic beam inside the cryostat and recaptured in front of the atom-chip by a mirror-MOT. After 2~s of loading, the atoms are compressed in a MOT whose magnetic field is created by the U-wire (Fig.~\ref{fig:approach}) and an external bias field. They are then cooled in an optical molasses and optically pumped to the $|F=2,m_F=2\rangle$ Zeeman sublevel of the ground state. The cloud is finally transferred, 460~\textmu m away from the surface, to an Ioffe-Pritchard trap produced by a current $I_Z=1.32$~A in the Z-wire (width 40~\textmu m) and a $\bm{B}=(B_x,B_y,B_z)=(0,0,4.4)$~Gauss bias field.

In the Ioffe trap, the atomic cloud is compressed by raising the magnetic bias field to $(11.2,0,26.2)$~Gauss in 20~ms. This increases the elastic collision rate and leads to a fast thermalisation of the sample. We perfom evaporative cooling for 5~s using an r.f. field radiated by another wire on the chip (not shown in Fig.~\ref{fig:approach}). The temperature after evaporative cooling  is set by the final value of the r.f. frequency ramp. This setting results from a compromise between the total number of atoms and the size of the cloud which must be smaller than the distance to the surface. We do not cool down to quantum degeneracy in order to avoid three-body collision losses. Atomic cloud temperatures are always measured after an adiabatic decompression of the trap to $\bm{B}=(11.2,0,8.4)$~Gauss~\cite{ENS_BECSUPRA08}. They range from 3 to 25~\textmu K.

The atomic cloud position is controlled by the $y$ and $z$ components of the bias field. The bias field along $x$ is set to $B_x=$~3.0~Gauss. Our choice of the bias field $B_x$ corresponds to a minimum of the technical magnetic-field noise spectrum in our apparatus and is compatible with a large trap depth at large distances. Neglecting the influence of the Z-wire current on $B_x$, the spin-flip Zeeman frequency at the bottom of the trap is expected to be $2.1$~MHz. 
By Zeeman spectroscopy, we have measured the actual frequency $\omega / 2\pi = 2.20(15)$~MHz. The uncertainty on $\omega$ includes the spread of the Zeeman frequency due to the Z-wire field for all atomic cloud positions of this experiment. We have checked that the measured lifetimes do not depend on $\omega$ within this uncertainty range. 

The atomic cloud is imaged by the absorption of a probe beam in the $xy$ plane at an angle of $11\,^{\circ}$ with respect to the $x$ axis. We thus view simultaneously the cloud and its reflection in the gold surface. The distance between these two images provides an accurate determination of the distance $d$ between the cloud and the surface. However, for $d\leq 50$~\textmu m, the two images merge because of the elongated shape of the atomic cloud along the $x$ direction. The distance $d$ is then evaluated by a linear extrapolation of its variation with the control parameters ($B_y$ and $B_z$), with large error bars at short distances. Obviously, when $d$ is smaller than the transverse size of the atomic cloud, additional losses occur due to the adsorption on the chip surface. For the lowest cloud temperature, $T=3$~\textmu K, this happens for $d\lesssim15$~\textmu m. Let us note also that the scattering of the incoming laser on the edge of the Z-wire is visible in the absorption image, giving us access to the $z$ coordinate of the cloud (Fig~\ref{fig:approach}). 

At the end of the evaporation, the atoms are at 60~\textmu m from the surface on the $y$ axis. By changing $B_z$ while keeping $B_y=0$, we can move the cloud within 400 ms adiabatically along the $y$ axis (figure \ref{fig:approach}). For this ``direct approach'', the distance $d$ to the surface is equal to the distance $r$ to the current-carrying Z-wire. The loss rate due to the technical current fluctuations in the Z-wire obviously depends on $r$. In order to measure independently the effect of the gold layer, which depends only on $d$, we also use a ``side approach'' in which $r$ is nearly constant~\cite{TR_HENKELSCHMIEDMAYER_SUBSURFACE_05}. Starting from the initial conditions, during 400 ms $B_z$ is decreased and $B_y$ increased so that $B_z+B_y$ is constant. The distance $d$ changes while $r$ is always greater than 160~\textmu m. We approach the surface on the side of the Z-wire opposite to the U-wire. In this way, we avoid the trapping potential perturbation due to the Meissner effect in the superconducting U-wire~\cite{TR_FORTAGH_MEISSNERTHEO_08}.

The final conditions at the end of the approach sequence are maintained for a variable duration $t$ before taking an image of the cloud, used to evaluate the number of atoms remaining in the trap. By repeating the experiment for different trapping times, we measure the trapping lifetime.

\section{Results}
\label{sec:observations}

\subsection{Direct approach measurements}
\label{sec:directapproach}

\begin{figure}
\resizebox{1.05\columnwidth}{!}{
\centerline{\includegraphics{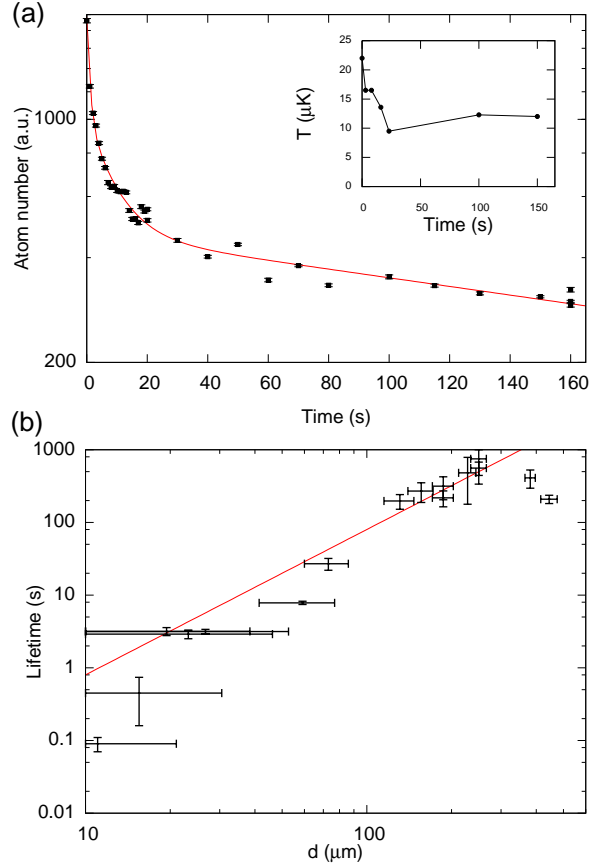}}
}
\caption{(a) Number of atoms as a function of the trapping time $t$ for a cloud at a distance $d=250$~\textmu m from the trapping wire. The solid line is a bi-exponential fit leading to decay times of 8.90(15)~s and 350(70)~s. The inset shows the temperature of the cloud as a function of time for the same conditions. The temperature decreases during the rapid decay of the atom number and then remains constant, supporting our assumption that evaporation takes place during this initial period. (b) Trap lifetime $\tau$ as a function of the distance $d$ in the case of the direct approach of Fig.~\ref{fig:approach}. The solid line $\tau \varpropto d^2$ is a theoretical fit, in good agreement with the measurements for $d$ between 20 and 300~\textmu m.}
\label{fig:directresults}
\end{figure}

We first describe the lifetime measurements in the case of the direct approach, when the distance to the surface, $d$, and the distance to the Z-wire, $r$, are equal. Fig.~\ref{fig:directresults}(a) presents a typical measurement of the number of atoms as a function of the trapping time $t$ for a distance $d=250$~\textmu m, together with a bi-exponential fit. Initially, we observe a rapid decay attributed to evaporation of the hottest atoms (time constant 8.90\,(15)~s). This interpretation is confirmed by a simultaneous rapid decrease of the cloud temperature, shown in the inset of Fig.~\ref{fig:directresults}(a). For $t>30$~s, this rapid evaporation comes to an end, and we observe a very long lifetime $\tau=350(70)$~s.

We present in Fig.~\ref{fig:directresults}(b) a Log-Log plot of the lifetime $\tau$, measured after the fast initial evaporation, as a function of $d$. At large distances, we observe extremely long trapping lifetimes, in the 10 minutes range, comparable to the best results obtained with centimetre-sized superconducting coils in a cryogenic environment~\cite{TR_LIBBRECHTCRYOTRAP95}. Our results are significantly better than those presented in Ref.~\cite{ENS_CHIPSUPRA06}, where a lifetime $\tau = 115$~s at a distance $d=350$~\textmu m was reported. At the same distance we now obtain $\tau = 560(100)$~s. This improvement is due to better screening of the experiment from external r.f. radiation, to the use of low-noise current power supplies, as well as to a careful filtering of the d.c. currents in the wires connecting the experiment to the outside.  Note that the longest trapping times observed set a limit to the maximum residual $^4$He pressure in the experimental cell $P<7\times 10^{-12}$~mbar. 

The lifetime clearly decreases as the cloud is brought closer to the chip and to the wire. Over a large distance range, our observations follow a $\tau \varpropto d^2$ variation, as shown by the solid line in Fig.~\ref{fig:directresults}(b). The observed losses are much faster than those expected to be due to the vicinity of a normal or superconducting metallic surface~\cite{TR_HENKELMETALNOISE05,TR_REKDALSUPERCONDNOISE06}. We thus attribute these fast losses to the spin-flips induced by technical current noise in the Z-wire. The corresponding spin-flip rate towards untrapped magnetic sublevels is $\Gamma_{\mbox{\tiny tech}} = (\mu_B g_F)^2 \mathcal{B}^2 /\hbar^2$~\cite{TR_HENKEL_TECHNOISE_99},  where $\mu_B$ is the Bohr magneton, $g_F$ the Land\'e factor of the F=2 manifold and $\mathcal{B}$ is the spectral density of magnetic noise at the location of the atoms. It is related to the current noise spectral density in the trapping Z-wire. Assuming an infinitely thin wire, we have:
\begin{equation}\label{eq:lifetime_technical}
 \tau_{\mbox{\tiny tech}} = \left( \dfrac{h}{\mu_0 \mu_B g_F} \right) ^2 \dfrac{d^2}{\mathcal{I}^2},
\end{equation} 
where $\mathcal{I}$ is the current noise spectral density. The solid line of Fig.~\ref{fig:directresults}(b) is a fit of the experimental points for $20 < d< 300$~\textmu m. It uses Eq.~\ref{eq:lifetime_technical} and leads to $\mathcal{I} = 1.3(3)$~nA$\cdot$Hz$^{-1/2}$. Technical current noise is thus the dominant loss source in the case of the direct approach.

Let us note that, for small and large distances, atoms are lost significantly faster than expected by the previous analysis. In order to reach large distances, we lower the external field $B_z$ and accordingly reduce the depth of the trap, leading to additional losses. At short distances, the cloud size is of the order of $d$ and surface adsorption comes into play. This effect could be enhanced by the existence of screening supercurrents in the trapping wire (Meissner effect) which flatten the trapping potential for distances comparable with the width of the trapping wire~\cite{TR_FORTAGH_MEISSNERTHEO_08}.

\subsection{Side approach measurements}
\label{sec:sideapproach}

\begin{figure}
\resizebox{1.05\columnwidth}{!}{
\centerline{\includegraphics{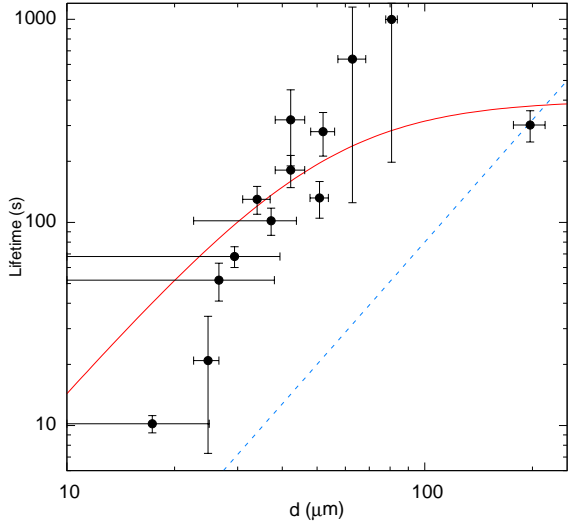}}
}
\caption{Lifetime measurements as a function of $d$ for the side approach. The dotted line presents, for reference, the lifetime observed for the direct approach. The solid line results from a theoretical model for the Johnson-Nyquist noise-induced losses, including the nearly constant contribution of the technical noise (see text). Theory is in good agreement with the observations.} \label{fig:sideresults}
\end{figure}

In order to reveal the effect of Johnson-Nyquist noise on the trapping lifetime, we use the side approach. In this case, the distance from the wire $r$ remains nearly constant (it varies between 160 and 240~\textmu m). Using the results of the previous section, the technical-noise- limited lifetime is  $\tau_{\mbox{\tiny tech}}>200$~s. Fig.~\ref{fig:sideresults} presents the observed lifetimes as a function of the cloud-surface distance $d$ for the side approach. It shows that $\tau$ decreases with $d$, well below the limit set by the technical noise. We attribute these additional losses to the effect of Johnson-Nyquist noise in the gold layer. 

Our experimental results can be compared with theoretical predictions. The three relevant length scales are the distance $d$, the thickness $h$ of the metallic film, and its skin-depth $\delta=\sqrt{2/(\mu_0 \sigma \omega)}$, which characterizes the dissipation in the material at the Zeeman transition frequency $\omega$ ($\sigma(\omega)$ is the conductivity of the metal). The trapping time can be calculated in the asymptotic regime corresponding to our experimental parameters where $h \ll \delta \lesssim d$~\cite{TR_HINDS_LIFETIMECONSUPERCON_05,TR_VARPULA_LIFETIMETHINMETAL_84,TR_HENKELMETALNOISE05}:
\begin{equation}\label{eq:varpula}
 \tau \approx \dfrac{\tau^0}{n_{th}+1} \dfrac{32}{9} \left( \dfrac{\omega}{c} \right)^3 \dfrac{d^2 \delta^2}{h} \left[ 1 + \left( \dfrac{4dh}{\pi^2 \delta^2}\right)^2 \right]
\end{equation} 
where $n_{th}$ is the mean thermal photon number at the transition frequency $\omega$ and $\tau^0$ is the lifetime in free space at $T=0$ given by the Wigner-Weisskopf formula. 
In the case of our experiment, $h =$ 200 nm, $\omega=2\pi\times 2.2$~MHz, $n_{th} = 4.0\times 10^{4}$ and $\tau^0=1.9 \times 10^{23}$~s.


In order to compare the theoretical prediction with the measurements, we must also take into account the nearly constant loss rate induced by technical current fluctuations. The solid line of Fig.~\ref{fig:sideresults} is a fit of the experimental data based on this approach, keeping as free parameters the  technical noise lifetime $\tau_{\mbox{\tiny tech, fit}}$ and the gold layer conductivity $\sigma$. We get $\tau_{\mbox{\tiny tech, fit}}=400\pm 200$~s, consistent with the results of the previous section. 
The conductivity is found to be $\sigma=6.7\pm 4.0 \times 10^9$~($\Omega$m)$^{-1}$, in excellent agreement with standard values for thin films of gold at 4.2~K ($\sigma=5 \times 10^9$~($\Omega$m)$^{-1}$~\cite{TR_LIDE_GOLDRES_1997}). It corresponds to a skin-depth $\delta=4.1$~\textmu m. We are thus well within the asymptotic limits of  Eq.~(\ref{eq:varpula}). 

\section{Conclusion and perspectives}

We have observed extremely long trapping lifetimes above a superconducting atom-chip in a cryogenic environment. We get lifetimes in the 10 minutes range at large distances (350 \textmu m) from the chip surface. This is a considerable improvement over previous experiments \cite{ENS_CHIPSUPRA06}, which is due to largely improved screening of magnetic field technical noise.

A careful control of the position of the trapped atomic cloud above the chip surface allowed us to explore the trap loss mechanisms when the atoms are brought close to the chip. At short distances to the trapping wire, the current-noise induced losses are dominant. Far from the wire, but at short distances from the thin gold layer covering the chip, we observe the effect of Johnson-Nyquist noise. The magnitude of this noise is in excellent agreement with the theoretical predictions. This effect sets a fundamental limit to the maximum achievable lifetime above a normal conducting surface.

The current-noise-induced losses could be considerably reduced by using permanent superconducting currents as realized in Ref.~\cite{TR_SHIMIZUPERMANENTCHIP07}, since in that case the technical noise coming from the external power supplies could be suppressed. The Johnson-Nyquist noise limit could be overcome if the atoms are brought close to a superconducting surface. In our experiment, it requires a new chip generation, where the atoms can approach a bare superconducting surface, which is not covered by a gold layer as it is the case for these measurements. In this case, however, additional loss mechanisms, due to the presence of Meissner-effect screening currents ~\cite{TR_FORTAGH_MEISSNERTHEO_08} could come into play. We have preliminary observations indicating that this effect is a limitation, when the atoms come very close to the trapping Z-wire in the direct approach. This limitation could be avoided by using a trapping wire whose width is much smaller than its distance to the trapped atoms. This would lead to long lifetimes and tight trapping potentials, promising for the coherent manipulation of atomic motion.

\section*{Acknowledgements}

This work was supported by the Agence Nationale pour la
Recherche (ANR), by the Japan Science and Technology Agency (JST), and by the
EU under the IP project SCALA. A.L. acknowledges support by an EU Marie Curie fellowship.

\bibliographystyle{epj}

\end{document}